# Integration of atomically thin layers of transition metal dichalcogenides into high-Q, monolithic Bragg-cavities - an experimental platform for the enhancement of optical interaction in 2D-materials

**HEIKO KNOPF,**[1,2,*] **NILS LUNDT,**[3] **TOBIAS BUCHER,**[1] **SVEN HÖFLING,**[3,4] **SEFAATTIN TONGAY,**[5] **TAKASHI TANIGUCHI,**[6] **KENJI WATANABE,**[6] **ISABELLE STAUDE,**[1] **ULRIKE SCHULZ,**[2] **CHRISTIAN SCHNEIDER**[3] **AND FALK EILENBERGER**[1,2,7]

[1]*Institute of Applied Physics, Abbe Center of Photonics, Friedrich Schiller University, Albert-Einstein-Straße 15, 07745 Jena, Germany*
[2]*Fraunhofer Institute of Applied Optics and Precision Engineering IOF, Center of Excellence in Photonics, Albert-Einstein-Straße 7, 07745 Jena, Germany*
[3]*Department of Technical Physics, Julius-Maximilians-Universität, Am Hubland, 97074 Würzburg, Germany*
[4]*SUPA, School of Physics and Astronomy, University of St. Andrews, St. Andrews KY169SS, UK*
[5]*Arizona State University, Tempe, Arizona 85287 USA*
[6]*National Institute for Materials Science, Tsukuba, Ibaraki 305-0044, Japan*
[7]*Max Planck School of Photonics, Part of the Max-Planck-Gesellschaft zur Förderung der Wissenschaften e.V., Albert-Einstein-Straße 7, 07745 Jena, Germany*

*\* heiko.knopf@uni-jena.de*

**Abstract:** We demonstrate a new approach to integrate single layer MoSe$_2$ and WSe$_2$ flakes into monolithic all-dielectric planar high-quality micro-cavities. These distributed-Bragg-reflector (DBR) cavities may e.g. be tuned to match the exciton resonance of the 2D-materials. They are highly robust and compatible with cryogenic and room-temperature operation. The integration is achieved by a customized ion-assisted physical vapor deposition technique, which does not degrade the optical properties of the 2D-materials. The monolithic 2D-resonator is shown to have a high Q-factor in excess of 4500. We use photoluminescence (PL) experiments to demonstrate that the coating procedure with an SiO$_2$ coating on a prepared surface does not significantly alter the electrooptical properties of the 2D-materials. Moreover, we observe a resonance induced modification of the PL-spectrum for the DBR embedded flake. Our system thus represents a versatile platform to resonantly enhance and tailor light-matter-interaction in 2D-materials. The gentle processing conditions would also allow the integration of other sensitive materials into these highly resonant structures.

## 1. Introduction

The discovery of graphene, with its unique characteristics, inspired a plethora of research activities in the field of monolayer materials [<akinwande2017>, <zhang2016>]. Atomically thin 2D-monolayers, in strong dissimilarity to their bulk counterparts [<lee2008>], can provide extraordinary characteristics, such as superconductivity [<saito2016>], better absorbance or higher transmission of light, altered bandgaps [<splendiani2010>] or extreme hardness [<akinwande2017>] to name a few.

Transition metal dichalcogenides (TMDCs) are semiconducting 2D-materials with direct bandgaps in the visible range from 1.0 to 2.5 eV. These consist of a layer of transition metals such as W or Mo sandwiched between two chalcogen layers, i.e. S, Se, or Te layers.

Monolayer TMDCs exhibit peculiar optical effects, which are related to the confinement of electronic motion in a 2D plane and the absence of dielectric screening, as well as to their crystal symmetry. The absorption of photons with energy above the bandgap in TMDCs causes the generation of hot electrons [<cui2014>], which swiftly form bound electron-hole pairs, termed excitons. Excitons in TMDCs are highly stable with binding energies in the range of hundreds of meV [<chernikov2014>]. Both the linear and nonlinear electronic [<zhu2015>, <zhang2017>] and optical [<saleh2018>, <tonndorf2015>] properties of TMDCs are strongly affected by these excitons. Due to their stability and robustness [<palummo2015>, <ruppert2017>], TMDCs are ideal candidates for exciton experiments. They exhibit non-linear properties [<saleh2018>, <nie2016>], making them interesting for experiments such as sum-frequency generation [<wang2015>-<janisch2014><janisch2014a>], but also for the generation of entangled photon pairs [<he2016>].

However, due to their single-layer nature, they are also highly susceptible to environmental parameters [<liu2015>], process conditions [<mcdonnell2016>], properties of the substrate material [<akinwande2017>, <lippert2017>], and substrate geometry [<kim2018>]. This makes experiments difficult to reproduce and highly dependent on laboratory conditions, which may be hard to control. The integration of TMDC layers in well-defined optical coatings and materials, such as glasses, would eliminate some of these issues and help establish TMDCs as a reproducible experimental platform.

Moreover, TMDCs are also interesting for the functionalization of classical optical materials. TMDC-loaded dielectrics may enable new classes of optical coatings. They may also operate as light emitters [<reed2015>, <aharonovich2016>]. The generated light has coherence properties which may be interpreted as lasing [<wu2014>, <waldherr2018>]. Naturally occurring [<tonndorf2015>, <srivastava2015>, <chakraborty2015>] or mechanically induced [<niehues2018>, <iff2018>] defect states can support the emission of single photons.

An application with highly challenging requirements for the integration of 2D-materials in optical systems comes in the form of strong coupling experiments [<lundt2016a>-<luo2018><shahnazaryan2017>]. Strong coupling refers to an exciton being coupled resonantly to an optical cavity of high quality and small modal volume. These excitons hybridize with the cavity mode and form so-called exciton-polaritons, the branches of which are separated by Rabi-splitting. Strong coupling can be observed if the dipole coupling strength, i.e. the product of the dipole moment $d$ of the exciton and the electric field $E$ at the position of the exciton, exceeds radiative and dissipative losses, e.g. photon leakage out of the cavity, represented by the cavities' quality factor (q-factor), and/or emitter dephasing [<savona1995>]. The q-factor is typically measured from spectral data as the ratio of the resonance wavelength and the line width of the cavity. As strong-coupling has already been demonstrated, is well understood and yet technically highly challenging, we find it to be a superb test-case to demonstrate the capability of our method to fabricate systems with a bandwidth and q-factor that is unpreceded for monolithic cavities.

For MoSe$_2$ it was shown [<lundt2016>] that monolithic distributed-Bragg-reflector-cavities (DBR-cavities) exhibit strict distinguishability [<liu2013>] of the Rabi-peaks for both cryogenic and room-temperature operation if a q-factor of $q \gg 1300$ [<lundt2016>] can be achieved. Although strong coupling was observed for lower q-factors [<liu2014a>], we use the predictions from [<lundt2016>] as a benchmark as it guarantees the strict distinguishability of the Rabi-peaks. It also opens a new path to high-quality, room-temperature polaritonic device architectures. Moreover, ion assisted PVD (IAD), employed here, generally imposes lower thermal loads than plasma-enhanced CVD (PECVD) [<liu2014a>], thus maximizing the selection of embeddable materials. IAD also has a larger set of materials to choose from, which can, for example, be used to implement higher refractive index contrasts. This also leads to a higher degree of flexibility and a broader range of applications for our method.

Recent results [<dhara2018>] underline the capabilities of PVD-techniques to fabricate systems for fundamental investigations in many-body polaritonics, which are only accessible to platforms with increased q-factors. Although the authors demonstrate a q-factor of $q = 600$, several questions remain open. These may be answered in systems with further increased q-factors; in accordance with the distinguishability-related benchmark $q \gg 1300$ derived above. Such systems can be attained with the IAD technique presented here.

Beyond strong coupling experiments, resonant structures have been utilized to modify the properties of spontaneous [<he2016>, <wu2014>] and stimulated emission [<tonndorf2015>, <wu2014>] in TMDCs. These experiments utilized open external cavities [<schwarz2015>, <dufferwiel2015>], metal-based cavities [<wang2016>] or nano resonators [<zheng2017>], all of which exhibit limited optical q-factors and/or large modal volumes. Other experiments that used photonic crystal resonators [<wu2014>, <fryett2017>, <ge2018>] exhibited a higher q-factor, but the 2D-material cannot be placed at the position of the peak field enhancement, thus the high q-factor cannot be exploited to the fullest.

In this work, we report on an ion-assisted physical vapor deposition process (IAD) with a temperature below 350 K used to embed TMDCs into a planar Fabry-Perot microcavity based on $SiO_2/TiO_2$ layerstacks (see Fig. 1). By using $TiO_2$ as a high refractive material and a high number of high-index-low-index-pairs (HL-pairs), we achieve access to increased q-factors and larger bandwidths. The process allows us to integrate exfoliated $MoSe_2$ and $WSe_2$ flakes with high-quality optical materials into monolithic, solid state layer systems with a high level of control on the material composition and thickness.

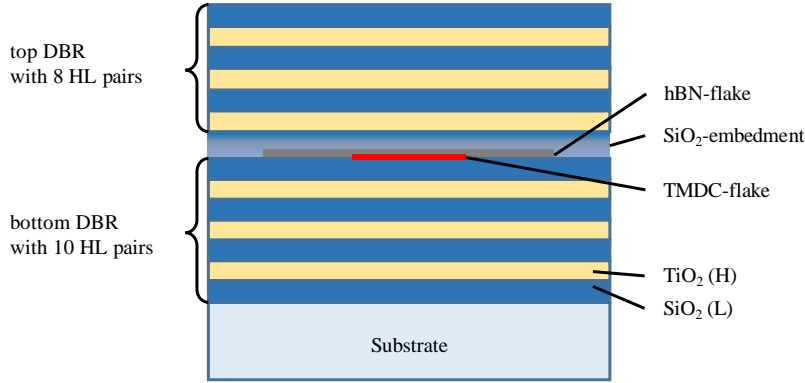

Fig. 1 Conceptual image of the embedded TMDC in planar Fabry-Perot microcavity

## 2. Methods

First, we determined the maximal q-factor of an unloaded Bragg cavity that can be achieved in our process. It is limited by the absorbance and the scattering of the coatings produced in the IAD process and by the maximal thickness of the layer stack, which can be fabricated without delamination. Both absorptive and scattering losses have been characterized for the IAD in prior works [<bennett1989>, <thielsch2002>]. The real wavelength dependent material parameters have been used for analytical calculation via OptiLayer, which we used to predict and optimize our structures [<tikhonravov2014-2018>].

The q-factor of the cavity depends on the transition bandwidth at the resonance position and hence increases with the number of high-index-low-index-layer-pairs used for both mirrors [<garmire2003>, <reichman2000>]. This can be seen in Fig. 2. Numerical calculations show that a q-factor of $q > 1300$ can be achieved with 7 HL-pairs on both sides of the cavity. For later ease of observation of strong coupling, we thus chose to pursue a design with 8 HL-pairs in the top mirror and 10 HL-pairs below. The number of HL-pairs at the bottom was increased to facilitate the emission of photoluminescence towards the top. Typically, layerstacks with a much larger thickness also tend to fail mechanically under thermal loading.

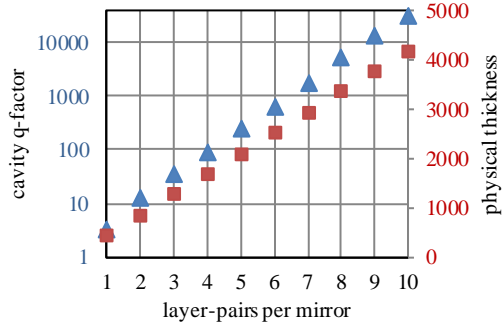

Fig. 2 Increase of the cavity q-factor of a symmetric cavity in dependency of the number of DBR-pairs for each mirror.

Both mirrors have been optimized for high reflectivity between 630 nm and 850 nm. The $TiO_2$ layer had a refractive index of $n_{TiO_2} = 2.284$ at 750 nm. The $SiO_2$ had a refractive index of $n_{SiO_2} = 1.455$ at 750 nm. The layers of both materials were tuned to $\lambda/4$-thickness resulting in 129.3 nm thick $SiO_2$-layers and 79.3 nm thick $TiO_2$-layers. Note that the design is limited to 300 nm for single $TiO_2$ layers to avoid detrimental influences from oversized polycrystalline growth, thus retaining smooth surfaces with low scattering, low absorbance, and high optical quality [<bennett1989>, <leprince-wang2000>]. The combined central $SiO_2$-spacer has an optical thickness of 375 nm to tune the resonance wavelength to $\lambda = 750$ nm, as confirmed by a pronounced dip in the reflection spectrum shown in Fig. 3(a). The calculated bandwidth of the resonance peak was 0.063 nm (full width at half maximum), equating into a q-factor of $q = 11900$.

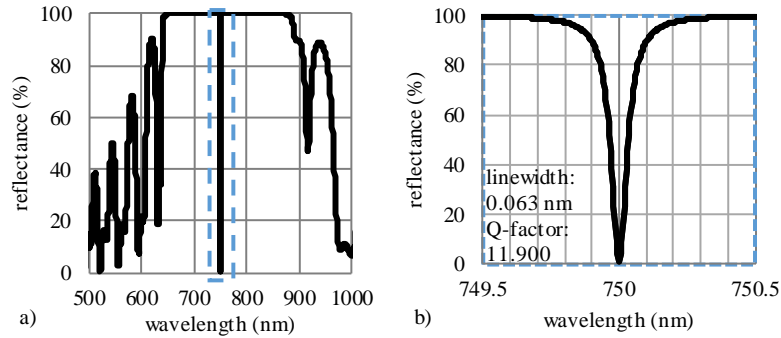

Fig. 3 a) Calculated reflectance curve for the optimized DBR-resonator with 10 HL layers at the bottom and 8 HL layer at the top. The resonance wavelength is 750 nm, illuminiation is at an angle of 0° from the top of the layer stack. b) Zoom to the resonance peak, marked with the blue dashed line in a).

Fig. 4 displays the calculated electric field intensity ($|E|^2$) for the realized 8/10-HL-stack design for on-resonance incident light at $\lambda = 750$ nm (see Fig.4(a)) and off-resonant excitation at $\lambda = 752$ nm (see Fig.4(b)) each at normal incidence. Because of the 8/10-HL-stack-design, the cavity is not symmetrical to the center. Nevertheless, the antinode position of the electromagnetic wave is at the middle of the spacer-area, coinciding with the position of the 2D-material, such that we can fully utilize the high q-factor.

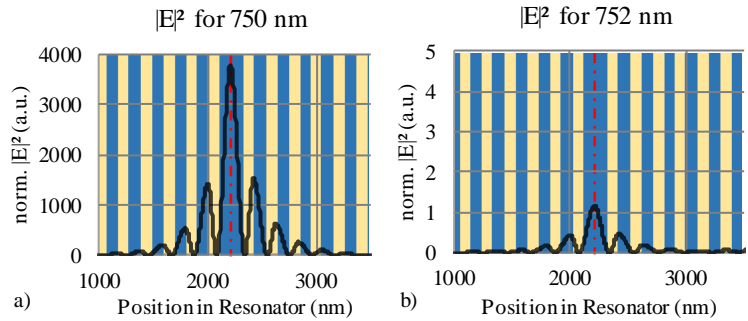

Fig. 4 a) Calculated normalized electric field intensity distribution for 750 nm input wavelength at an angle of 0°. The peak position is coincident with the position of the TMDC flake in the spacer area of the resonator (see schematic coating above the graph: blue=$SiO_2$, yellow=$TiO_2$, red=TMDC) b) Same as in a) but for non-resonant excitation at 752 nm wavelength. (schematic coating system was explained in Fig. 1)

The fabrication of the TMDC-loaded cavity was carried out in an ion-assisted deposition (IAD) process. Mechanically exfoliated $MoSe_2$ and $WSe_2$ monolayer flakes [<dean2010>] were placed on an optical base substrate, in our case a sputtered DBR made from 10 pairs of $SiO_2$ and $TiO_2$ layers, deposited on a quartz substrate. A few-layer boron nitride flake (hBN) with a thickness of about 10 nm was deposited on the 2D-flakes to protect the TMDC-flakes from influences caused by the subsequent coating process [<dean2010>]. The second cavity mirror consisting of eight $SiO_2$-$TiO_2$-pairs was deposited directly on the top via IAD preserving gentle coating conditions to comply with the weak van-der-Waals adhesion of the TMDC flakes.

The deposition procedure was performed in a physical vapor deposition plant Buhler SyrusPro1100 using a background pressure of about $10^{-5}$ mbar with a maximal process temperature of about 80 °C. An overview of the coating process is depicted in Fig. 5.

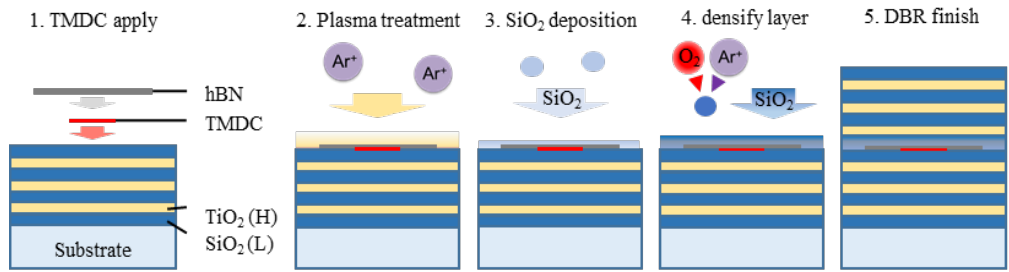

Fig. 5. Process steps for embedding 2D-monolayers into monolithical, optical DBR-cavites without damage to the TMDC flake.

The $SiO_2$ surface of the bottom DBR with the van-der-Waals-bound TMDC and hBN flakes on top was pretreated with an Ar-Ion plasma using 60 V Bias and 30 A discharge current with 10 sccm Ar for 7 seconds. This enhances the surface energy of the $SiO_2$-top layer by cracking OH-bonds, creating chemically active sites, to which subsequently deposited material may crosslink [<meyer1974>-<martinu2000><terpilowski2015><bhattacharya2005>]. The pretreatment provides an additional cleaning effect for the surface. Plasma exposure time and plasma energy have been determined from prior experiments to be sufficient to create a significant adhesion effect while maintaining a low dose to prevent delamination of the TMDC flake from the bottom mirror. No resulting increase in surface roughness was observed.

Next, we deactivated the plasma and coated the activated surface with $SiO_2$ evaporated by an electron beam using a low deposition rate of about 0.4 nm/s. This rate was chosen as the lowest reproducible deposition rate as its typical fluctuation is in the order of 0.2 nm/s. The first 10 nanometers of the $SiO_2$-layer were deposited without plasma assistance to prevent an overexposure of the surface while the coating is still thin and may not yet be fully coalesced.

The $SiO_2$ builds amorphous layers covering both the TMDC islands as well as the surrounding dielectric surface. An important parameter is the densification with argon and oxygen ions. The plasma leads to highly densified layers and increased refractive indices. It also induces compressive stress and reduces the tendency of the material to delaminate caused by tensile stress-induced cracking [<teixeira2001>]. This enhances the capability of the system to withstand large temperature differences and allows us to deposit more layers in a more reproducible manner. In a next step, the densification of the layer material was slowly increased by raising the Ar-flux and ion energy values typical for IAD [<schulz1996>]. For the outer layers of the DBR, a densification with up to 150 V bias voltage was used. Because of stoichiometry considerations, $O_2$ was added to the active plasma gas with 10 sccm gas flux for $SiO_2$-layers and 30 sccm for $TiO_2$.

For the fabrication of the cavity, it must be considered that both the TMDC as well as the hBN-flake contribute to the optical path in the spacer layer, and thus to the resonance wavelength. Their respective refractive indices have been taken from literature values [<li2014>-<song2010><zhang2017a>]. The thickness of the TMDC-layer of about 0.65 nm was based on findings in the literature [<morozov2015>], whereas the thickness of the hBN-flake of about 15 nm was measured with atomic force microscopy (AFM) (Bruker Dimension Edge AFM). The $SiO_2$ thickness of the spacer layer was reduced accordingly. The gradient refractive index of the less densified $SiO_2$-layer, as well as the thickness of the higher refractive $SiO_2$-sheath, was taken into account as well.

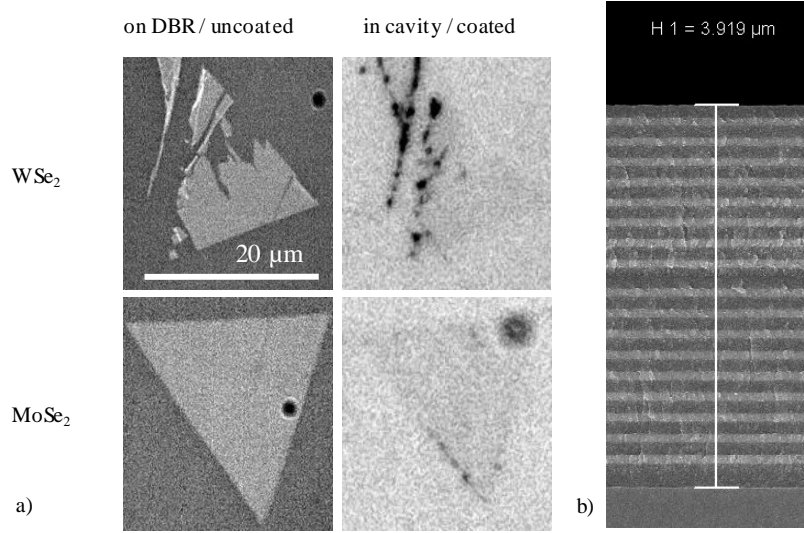

Fig. 6 a) Optical microscopy image of the 2D-materials without hBN from a preliminary run before (left) and after (right) treatment in the coating chamber. The integrity of the flakes is preserved. b) Scanning electron microscopy image of the deposited DBR-cavity from a preliminary run with the complete thickness of about 3.92 µm (just 9HL-pairs for bottom mirror for this specific sample). The layers have low individual and residual surface roughness, clear layer structure and a homogeneous spacer area in the middle

A microscopic top-view of a TMDC-loaded part before and after the coating process of the cavity is depicted in Fig. 6(a). While the contrast is greatly reduced due to the reflected light from the top-mirror, it still can be seen that the process does apparently not damage the TMDC-flakes. An SEM of the cavity cross section is depicted in Fig. 6(b). Due to the limited resolution of the SEM, the TMDC-layer cannot be observed directly.

The fabricated cavities have been subjected to cryostatic conditions and undergone multiple cooling-heating cycles between 5 K and 300 K without any signs of delamination or damage to the TMDCs.

## 3. Results

First, we verified in two steps if it is possible to coat the TMDCs with $SiO_2$ layers. A prior experiment on the adhesion properties was performed. TMDCs placed on the bottom mirror and covered with approximately 10 nm thick hBN sheets were treated with Ar-plasma. Contact-angle measurements showed sufficient increase of the surface energy providing the required precondition for adherent coating on these surfaces without delamination of the 2D-materials.

We then focused on the question of whether it is possible to preserve the structural and electrooptical properties of the 2D-material. We therefore coated 20 nm $SiO_2$ without plasma assistance directly on the 2D-flake. This second experiment was conducted to yield more specific information on the influence of embedding the TMDC on its photoluminescence (PL) properties. Room-temperature PL experiments were carried out with a 532 nm excitation laser providing 400 µW energy. The influence of the $SiO_2$ on the PL of the hBN covered $WSe_2$ - flakes is shown in Fig. 7(a). The increase of the layer thickness to 120 nm caused no shift of the PL peak positioned at 745 nm. A linewidth of roughly 40 meV could be achieved.

In Fig. 7(b), we present PL measurements of $MoSe_2$ with hBN-cover, both with 20 nm and 120 nm $SiO_2$ coating at room temperature and at 5 K. The linewidth is about 40 meV at room temperature and about 8 meV for excitonic and trionic resonances at 5 K. These results are comparable to former experiments by Lundt et.al. [<lundt2018>]. The prominent splitting in two peaks at cryogenic conditions indicates, that both excitonic and trionic oscillations are essentially unaffected by the application of the $SiO_2$ coating.

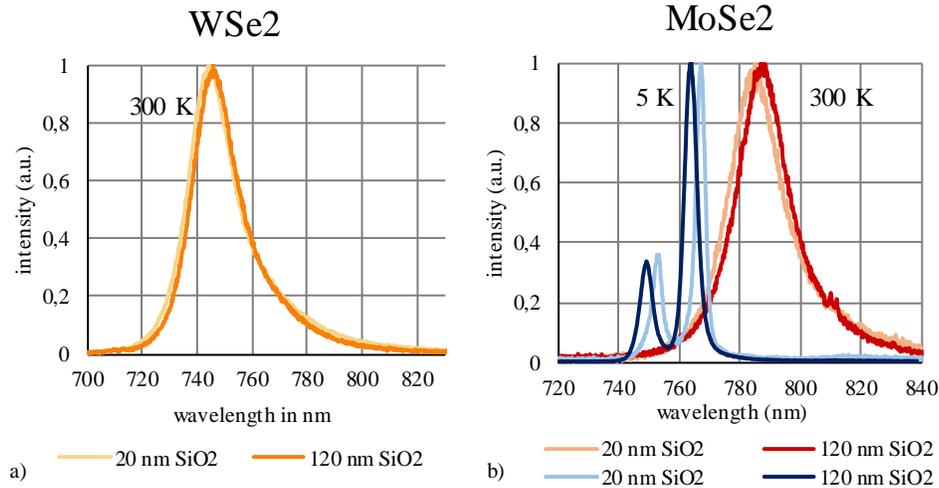

Fig. 7. PL intensity of $SiO_2$ covered layers. a) $WSe_2$-PL hBN covered flakes for 20 nm and 120 nm $SiO_2$ thickness b) $MoSe_2$-PL for 20 nm and 120 nm $SiO_2$ thickness at 300 K and 5 K

Next, we measured the optical reflectance of the deposited layer stacks using a standard UV/VIS spectrometer (Lambda 900 by Perkin Elmer), as well as a UV-NIR Micro-Spectrometer (USPM by Olympus). The subsequent morphological investigation of the encapsulation and the material distribution in the cavity was carried out with an optical stereo microscope by Leica systems and with an SEM-system Sigma by Carl Zeiss (Fig. 6(b)).

Prior to the complete embedding of the TMDC, we analyzed the optical performance on a bare DBR-mirror coated on a plane substrate to verify the validity of our DBR coating process. The reflection spectrum of the mirror and calculated design are depicted in Fig. 8(a). Both are in accordance, which proves that our coating process is operating as predicted.

Next, we fabricated a TMDC loaded DBR-cavity as discussed in the methods section. A measured reflection spectrum is depicted in Fig. 8(b). A reflectance spectrum of the DBR cavity at the resonance frequency is provided in Fig. 8(c). It shows the observed resonance at $\lambda = 749.3$ nm, which is roughly 0.1 % off the target value. This is consistent with typical variations of the coating process. A resonance bandwidth of $\Delta\lambda = 0.16$ nm was determined, which equates into a quality factor of $q = 4683$. This value is lower but quite in the magnitude of the design value of 11900. The difference between measurement and calculations may be attributed to defect spots in the layers, slight surface roughness and inhomogeneity's of refractive indices.

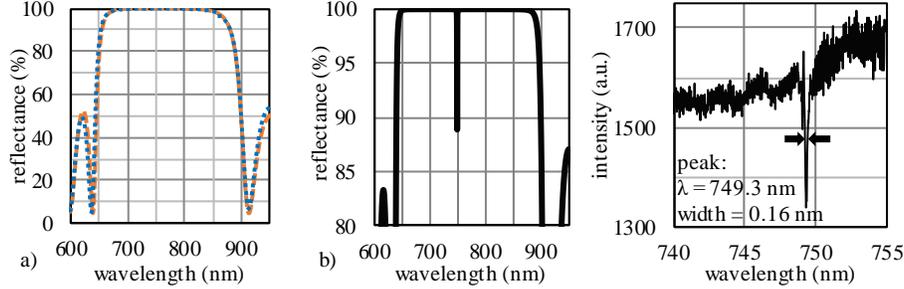

Fig. 8. a) Reflectance spectra of a bare DBR mirror calculated (red) and measured (blue) b) Measured broadband reflectance spectrum of a DBR-resonator cavity without TMDC c) Measured narrowband spectrum of reflection intensity around the cavity's resonance of TMDC loaded cavity at 749.3 nm with line width of 0.16 nm.

Then we verified, that the PL properties of the TMDC are unaffected by the DBR stack and will not produce any kind of background fluorescence, which would later negatively affect possible experiments at exciton wavelength.

To perform cross-sectional PL measurements of the DBR stack, a cross-sectional lamella was prepared from the cavity by means of Focused Ion Beam (FIB) milling using a FEI Helios NanoLab G3 UC. The lamella was attached to a TEM lift-out grid and thinned down to a final thickness of 200 nm with 30 kV Ga ions. No further low energy cleaning to remove amorphous layers or Ga ion contamination was performed. An SEM image of the lamella is depicted in Fig. 9(a).

Following, the lamella was transferred to a confocal laser-scanning microscope (PicoQuant MicroTime200). The microscope was used with a 40x/0.65NA objective corresponding to lateral resolution of about 1 µm with an excitation laser working at 532 nm with 80 MHz rep. rate and about 100 ps pulse length. The PL light was filtered with a 715 nm long pass filter. A measurement area of 40x40 µm was scanned with piezo positioning. The dataset of the measurement area was integrated along the lateral axis to receive a linescan, orientated perpendicular to the cavity system. The ensuing PL signal is superimposed on the SEM-image in Fig. 9(a). It shows two fluorescence peaks, one emanating from the expected location of the TDMC-flake, the other one from the substrate material at the bottom.

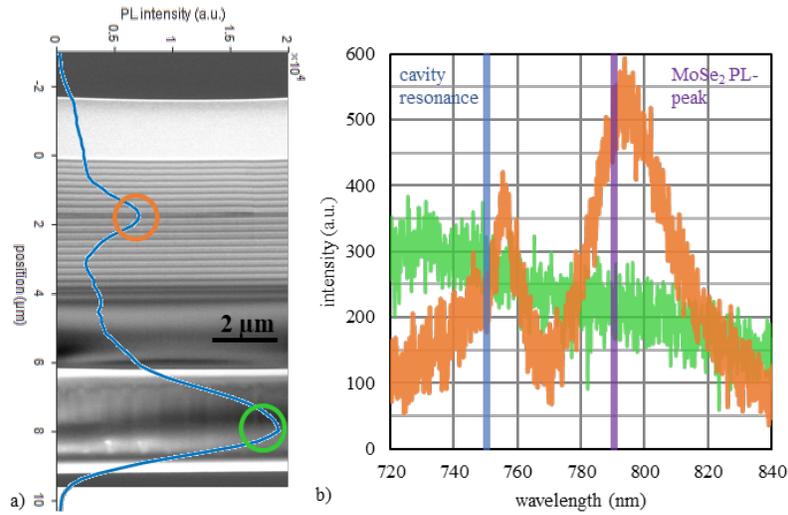

Fig. 9. a) FIB-lamella with superimposed PL intensity-cross section. Prominent peaks occur in the spacer-area (marked with orange circle) and the substrate-area (marked with green circle) b) Spectrum of the PL-light collected at the positions of the peaks marked in (a), colors of spectra match the colors of the circles. The resonance position of the unperturbed cavity (749.3 nm) and the position of the excitonic peak the PL of MoSe$_2$ (790 nm) are marked by the blue and purple vertical lines, respectively.

To further clarify the nature of these two PL peaks, we measured their spectra at the peak locations marked with the colored circles in Fig. 9(a). We used a Horiba spectrometer iHR320 with an integration time of 600 s operating at room temperature. The two spectra are shown in Fig. 9(b). The PL in the substrate exhibits a flat spectrum and is therefore not related to the MoSe$_2$ but rather to residual defect mediated autofluorescence of the substrate with a high contamination of Ga due to the lamella cutting process [<vaskin2018>]. The PL signal in the spacer area shows two spectral peaks. The first spectral peak occurs close to the resonance wavelength of the cavity at 755 nm. The small difference to the cavity wavelength may either be caused by self-bending of the membrane, or due to cracking and subsequent extension of the spacer-layer, which both start to occur during FIB-milling at the layer thickness of 200 nm. The second spectral peak occurs at 780-795 nm. While the latter peak clearly represents the characteristic A-exciton PL wavelength of MoSe$_2$ [<tonndorf2013>], we attribute the former to the action of the cavity. The PL peak is close to the position reported in the literature. Differences may be caused by the influence of the doped embedding material and from the strain induced by the FIB treatment. At this wavelength, the PL of the exciton is indeed enhanced by the cavity. Note that in this cross-sectional membrane, we cannot expect a high q-factor as it is only 200 nm thin and has a highly scattering surface.

These results show the presence of pristine, high-quality MoSe$_2$ in the cavity, the electronic properties of which have not been affected in a detrimental manner by the coating process. It also shows that its excitons do indeed couple to the cavity mode. WSe$_2$ would show similar results as suggested by the previous experiments.

## 4. Conclusion

We demonstrated a new approach to integrating single layer MoSe$_2$ and WSe$_2$ flakes into dielectric optical coatings and layer stacks. Our approach is based on a modified Ion Assisted Physical Vapor Deposition process. The gentle processing conditions allow us to integrate 2D-materials into optical coatings and layer stacks.

We selected the integration into monolithic, all-dielectric, high-q planar DBR-cavities as a benchmark for our process. This was selected for possible experiments on strong coupling and polaritronics, which require both high-quality 2D-materials as well as high-quality, small-volume resonators. The monolithic cavity could be realized without cracks, without damage to the TMDC, and with accurate reproduction of the theoretical layer-design. The ratio of the resonance wavelength and the line width of the resonance (Q-factor) of the cavity was higher than 4500 at 749.3 nm.

The presence of TMDC-material in the resonator, as well as its being unaffected by the coating process, was proven with photoluminescence measurements. For DBR-cavities, we observed photoluminescence from the $MoSe_2$ exciton at its fundamental wavelength and an enhancement of the PL emission at the slightly detuned cavity resonance. Our results suggest that the process presented in this work provides a viable platform for the study of strong coupling, polaritronics, and the enhancement of nonlinear-optical effects in 2D TMDC.

## 5. Acknowledgements

This work has been supported by the Fraunhofer-Gesellschaft zur Förderung der angewandten Forschung e.V.. We gratefully acknowledge the financial support by the German Federal Ministry of Education and Research via the funding "2D Nanomaterialien für die Nanoskopie der Zukunft" FKZ: 13XP5053A. Financial support from the Thuringian State Government within its Pro-Excellence initiative (ACP$^{2020}$) is gratefully acknowledged. The Würzburg group gratefully acknowledges financial support by the state of Bavaria. C.S. acknowledges support by the European Research Council within the project UnLiMIt-2D.